\documentclass[prd,aps,floats,twocolumn,showpacs,superscriptaddress,preprintnumbers,showkeys]{revtex4-1}

\usepackage[latin1]{inputenc}                    % Codepage latin1
\usepackage{graphicx}                            % Graphics package             
\usepackage{latexsym}                            % Load additional symbols
\usepackage{amsfonts}                            % Use AMS fonts
\usepackage{amssymb}                             % and symbols
\usepackage{amsmath}                             % and definitions
\usepackage[mathscr]{eucal}                      % Euler font symbols
\usepackage{dcolumn}                             % columns
\usepackage{theorem}                             % Allow use of theorems

\usepackage{subfigure,feynmf}

\begin{document}

\preprint{
\vbox{
\hbox{ADP-13-11/T831}
}}

\title[Charge symmetry breaking from a chiral extrapolation of moments of quark distribution functions]{Charge symmetry breaking from a chiral extrapolation of moments of quark distribution functions}
\author{P.E.~Shanahan}\affiliation{ARC Centre of Excellence in Particle Physics at the Terascale and CSSM, School of Chemistry and Physics, University of Adelaide, Adelaide SA 5005, Australia}
\author{A.W.~Thomas}\affiliation{ARC Centre of Excellence in Particle Physics at the Terascale and CSSM, School of Chemistry and Physics, University of Adelaide, Adelaide SA 5005, Australia}
\author{R.D.~Young}\affiliation{ARC Centre of Excellence in Particle Physics at the Terascale and CSSM, School of Chemistry and Physics, University of Adelaide, Adelaide SA 5005, Australia}

\begin{abstract}
We present a determination, from lattice QCD, of charge symmetry violation in the spin-independent and spin-dependent parton distribution functions of the nucleon. This is done by chirally extrapolating recent QCDSF/UKQCD Collaboration lattice simulations of the first several Mellin moments of the parton distribution functions of octet baryons to the physical point. We find small chiral corrections for the polarized moments, while the corrections are quantitatively significant in the unpolarized case.
\end{abstract}



\maketitle

%--------------------------------------------------------------------------
%
\section{Introduction}

Charge symmetry refers to the equivalence of $u$ quarks in the proton and $d$ quarks in the neutron, and vice-versa. Precisely, it is the invariance of the strong interaction under a rotation of $180^\circ$ about the `2'-axis in isospin space.
%Formally expressed in terms of quark distributions, this symmetry implies
%
%\begin{align}
%u^p(x,Q^2)&=d^n(x,Q^2), & d^p(x,Q^2)&= u^n(x,Q^2).
%\end{align}
%
%
At low energies, charge symmetry is obeyed to a precision of order 1\%~\cite{Miller:2006tv}. It would be natural to expect that partonic charge symmetry should hold to a similar extent. Traditionally, this expectation has been applied to parton phenomenology~\cite{Londergan:1998ai,Londergan:2009kj}, and the assumption of good charge symmetry has been used to reduce the number of independent quark distribution functions by a factor of two. 

Recently, charge symmetry violating (CSV) effects have been included in phenomenological parton distribution functions for the first time~\cite{Martin:2003sk}, and theoretical estimates of the size of such effects have been made~\cite{Rodionov:1994cg,Sather:1991je}. Experimental upper limits on partonic CSV are in the range 5-10\%~\cite{Londergan:1998ai,Londergan:2009kj}. CSV of this magnitude would produce important effects in tests of physics beyond the standard model, for example in neutrino deep inelastic scattering experiments~\cite{Londergan:2003pq}.

The first two Mellin moments of the spin-dependent quark distribution functions of the octet baryons, and the second spin-independent Mellin moment, have recently been determined from $N_f=2+1$ lattice simulations by the QCDSF/UKQCD Collaboration~\cite{Horsley:2010th,*Cloet:2012db,Bietenholz:2011qq}.
The first analysis of this lattice data used a linear flavour expansion about the simulation SU(3) symmetric point to extract values for the charge symmetry violating distributions~\cite{Horsley:2010th,Cloet:2012db}. Using chiral extrapolation formulae for the Mellin moments of quark distribution functions~\cite{Arndt:2001ye,Diehl:2006js,Dorati:2007bk,Chen:2001pva,Burkardt:2012hk,Chen:2001eg,Detmold:2005pt}, recently extended to include CSV effects~\cite{Shanahan:2012}, we improve on the original analysis by extrapolating the lattice results to the physical point. We find that chiral physics generates small corrections to the parton CSV terms.

%This paper is structured as follows:.....(?)

\section{Method}
\label{sec:method}

In terms of quark distributions, charge symmetry implies
\begin{align}
u^p(x,Q^2)&=d^n(x,Q^2), & d^p(x,Q^2)&= u^n(x,Q^2),
\end{align}
with analogous relations for the antiquark distributions.
A measure of the size of the violation of charge symmetry is given by the `CSV parton distributions', defined in terms of the Mellin moments as
\begin{align}
\label{eq:deltau}
\delta u^{m\pm} &= \int_0^1 dx x^m (u^{p\pm}(x)-d^{n\pm}(x))  \\
& = \langle x^m \rangle_u^{p\pm} - \langle x^m \rangle_d^{n\pm} \label{eq:deltad}
\intertext{and}
\delta d^{m\pm} &= \int_0^1 dx x^m (d^{p\pm}(x) - u^{n\pm}(x))  \\
&= \langle x^m \rangle_d^{p\pm}- \langle x^m \rangle_u^{n\pm}
\end{align}
for the spin-independent distributions, with analogous expressions for the spin-dependent case. Here, the plus (minus) superscripts indicate C-even (C-odd) distributions:
\begin{equation}
q^{\pm}(x) = q(x) \pm \overline{q}(x).
\end{equation}

The Mellin moments accessible to lattice simulations alternate between C-even and -odd moments with increasing $m$; the
 ($m-1$)th spin-independent (SI) and $m$th spin-dependent (SD) lattice moments are defined as
\begin{align}
\langle x ^{m-1} \rangle^B_q = & \int^1_0 dx x^{m-1} ( q^B(x) + (-1)^m \overline{q}^B(x)),\\
\langle x ^{m} \rangle^B_{\Delta q} = & \int^1_0 dx x^{m} (\Delta q^B(x) + (-1)^m \Delta \overline{q}^B(x)).
\end{align}
 
Recent lattice simulations from the QCDSF/UKQCD Collaboration~\cite{Horsley:2010th,Cloet:2012db} give results for the first two SD and first SI lattice moments.
As these $N_f=2+1$ lattice simulations use degenerate light quarks, the CSV terms cannot be directly evaluated from the simulation results using Eqs.~(\ref{eq:deltau}) and~(\ref{eq:deltad}) (as this would give zero in each case). The problem can, however, be approached indirectly.

The original analysis of the QCDSF/UKQCD Collaboration lattice data used a linear flavour expansion at the simulation SU(3) symmetric point to estimate the CSV terms~\cite{Horsley:2010th,Cloet:2012db}. That is, the CSV terms were expressed in terms of hyperon moments as 
\begin{equation}
\delta u = m_\delta \left( - \frac{\partial \langle x \rangle_u^p}{\partial m_u}+ \frac{\partial \langle x \rangle^p_u}{\partial m_d} \right) + \mathcal{O}(m_\delta^2),
\end{equation}
where $m_\delta = (m_d-m_u)$, and $\delta d$ may be written in a similar way. The equivalence of the $u$ and $d$ quarks in the simulations, i.e., that $\partial \langle x \rangle^n_d / \partial m_d = \partial \langle x \rangle_u^p/\partial m_u$ and $\partial \langle x \rangle_d^n/\partial m_u = \partial \langle x \rangle_u^p/\partial m_d$, has been used to simplify the expression.

Near the SU(3) symmetric point, the strange quark is considered as a `heavy light quark', so that
\begin{align}
\frac{\partial \langle x \rangle^p_u}{\partial m_u} &\approx \frac{\langle x \rangle_s^{\Xi^0}-\langle x \rangle^p_u}{m_s-m_l}, &\frac{\partial \langle x \rangle^p_u}{\partial m_d} & \approx \frac{\langle x \rangle^{\Sigma^+}_u - \langle x \rangle^p_u}{m_s-m_l}, \\
\frac{\partial \langle x \rangle^p_d}{\partial m_u} &\approx \frac{\langle x \rangle_u^{\Xi^0}-\langle x \rangle^p_d}{m_s-m_l}, &\frac{\partial \langle x \rangle^p_d}{\partial m_d} & \approx \frac{\langle x \rangle^{\Sigma^+}_s - \langle x \rangle^p_d}{m_s-m_l}.
\end{align}
Rearranging, the CSV momentum fractions can be written as~\footnote{\label{theFoot} In references~\cite{Horsley:2010th,Cloet:2012db}, the factor of $(1/2)$ appearing at the beginning of the following equations was erroneously omitted. As a result, the values quoted for the CSV terms were too large by a factor of two. Corrected results are given in the first column of Table~\ref{tab:results2}.}
\begin{align}
\label{eq:lineq}
\frac{\delta u}{\langle x \rangle^p_{u-d}}&= \frac{1}{2}\frac{m_\delta}{\overline{m}_q} \frac{(\langle x \rangle_u^{\Sigma^+}-\langle x \rangle_s^{\Xi^0})/\langle x \rangle^p_{u-d}}{(m_K^2-m_\pi^2)/X_\pi^2},\\ \label{eq:lineq2}
\frac{\delta d}{\langle x \rangle^p_{u-d}}&=\frac{1}{2}\frac{m_\delta}{\overline{m}_q} \frac{(\langle x \rangle_s^{\Sigma^+}-\langle x \rangle_u^{\Xi^0})/\langle x \rangle^p_{u-d}}{(m_K^2-m_\pi^2)/X_\pi^2},
\end{align}
where $\overline{m}_q=(2m_l+m_s)/3$ and $X_\pi^2 = (2m_K^2+m_\pi^2)/3$.
Similar expressions hold for the spin-dependent CSV moments. This method allows an estimate of CSV at the SU(3) symmetric point. 

To evaluate the CSV terms at the physical point, we perform a chiral extrapolation of the lattice data for the quark moments~\cite{Shanahan:2012}. As the isospin-averaged and -broken expressions for the Mellin moments as functions of pseudoscalar or quark mass have the same unknown parameters, a fit to the available isospin-averaged lattice results allows the CSV terms to be evaluated from the isospin-broken expressions using Eqs.~(\ref{eq:deltau}) and (\ref{eq:deltad}) - a technique also used in~\cite{Shanahan:2012wa}. These expressions can be evaluated at any pseudoscalar masses, in particular at the physical point.

\section{Extrapolation of lattice data}
\label{sec:latticefits}

\subsection{Fit to isospin-averaged lattice data}
\label{subsec:isoavfit}

In previous work, we described an isospin-averaged chiral perturbation theory fit to QCDSF/UKQCD Collaboration lattice data for the first few Mellin moments of quark distributions~\cite{Horsley:2010th,Cloet:2012db}. Complete details of the fit formulae, fit parameters and method are given in Ref.~\cite{Shanahan:2012}. 

In brief, chiral perturbation theory expansions, described in Ref.~\cite{Shanahan:2012}, were fit to QCDSF/UKQCD Collaboration lattice data for the first spin-independent and zeroth and first spin-dependent moments. The fit functions include loop corrections and counterterms to leading non-analytic order. In particular, the effect of chiral loops with both octet and decuplet baryon intermediate states, as well as, for the spin-dependent moments, loops involving a transition between octet and decuplet baryons, are included. Tadpole diagrams and terms representing wavefunction renormalization are also considered. 

The finite-range regularization scheme (FRR) is used to regularize the loop integrals. This technique, discussed further in Refs.~\cite{Leinweber:2003dg,Young:2002ib,Young:2002cj}, involves the introduction of a mass scale $\Lambda$ through a regulator $u(k)$ inserted into each integral expression. $\Lambda$ is related to the scale beyond which a formal expansion in powers of the Goldstone boson masses breaks down (this scale is typically $\sim \Lambda/3$ for a dipole). For this analysis, a dipole regulator $u(k)=\left(\frac{\Lambda^2}{\Lambda^2+k^2}\right)^2$ and a regulator mass $\Lambda = 1$~GeV are chosen. This is based on a comparison of the nucleon's axial and induced pseudoscalar form factors~\cite{Guichon:1982zk} and the value of $\Lambda$ deduced from a lattice analysis of nucleon magnetic moments~\cite{Hall:2012pk}. All results are insensitive to this choice; choosing different regulator forms, for example monopole, Gaussian or sharp cutoff, and allowing $\Lambda$ to vary by $\pm 20 \%$ does not change the results of the analysis within the quoted uncertainties.

Within the FRR framework, expressions for loops with octet intermediate states involve the integral:
\begin{equation}
\label{eq:octint}
J(m^2) = \frac{4}{3} \int_0^\infty dk \frac{k^4u^2(k)}{(\sqrt{k^2+m^2})^3}
\end{equation}
with the finite-range regulator $u(k)$ inserted. The normalization of $J(m^2)$ has been defined so that the non-analytic part matches the common form of dimensionally regularized results, as $J(m^2) \underset{DR}{\rightarrow} m^2 \textrm{ln}(m^2/\mu^2)$.
Loops with decuplet intermediate states may be written in an analogous way, in terms of
\begin{align}
J_1(m^2,\delta) & = \frac{4}{3} \int_0^\infty dk \frac{k^4u^2(k)}{(\sqrt{k^2+m^2})^2(\sqrt{k^2+m^2}+\delta)}\\
\intertext{and}
J_2(m^2,\delta) & = \frac{4}{3} \int_0^\infty dk \frac{k^4u^2(k)}{(\sqrt{k^2+m^2})(\sqrt{k^2+m^2}+\delta)^2},
\end{align}
which describe loops with one and two decuplet propagators respectively. The tadpole contributions are written in terms of
\begin{equation}
\label{eq:tadeq}
J_T(m^2)=4 \int_0^\infty dk \frac{k^2u^2(k)}{\sqrt{k^2+m^2}}
\end{equation}
which has the same non-analytic structure as $J$, i.e., $J_T(m^2) \underset{DR}{\rightarrow} m^2 \textrm{ln}(m^2/\mu^2)$. To make comparison with DR expressions clear and to avoid absorbing loop terms into known parameters such as $F$ and $D$, constant terms are subtracted by the integral replacement
\begin{equation}
\mathcal{I}(m_\phi) \rightarrow \widetilde{\mathcal{I}}(m_\phi)=\left[\mathcal{I}(m_\phi)-\mathcal{I}(m_\phi=0)\right],
\end{equation}
where $\mathcal{I}$ represents any of the integrals in Eqs.~(\ref{eq:octint})--(\ref{eq:tadeq}).

The fit to the lattice results is performed by minimizing the sum of $\chi^2$ for each set of moments. There are 24 lattice data points available for each moment considered~\cite{Horsley:2010th,Cloet:2012db,private}. The fit parameters, discussed in detail in Ref.~\cite{Shanahan:2012} and listed in Appendix~\ref{app:CSVexpand}, are different for each moment. For the zeroth spin-dependent moment there are eight free parameters, while both the first spin-dependent and first spin-independent moment have nine fit parameters.

Figures~\ref{fig:SD0}, \ref{fig:SD1}, \ref{fig:SI1}, taken from Ref.~\cite{Shanahan:2012} and located in Appendix~\ref{app:figs}, show the quality of fit for each moment. Here $X_\pi = \sqrt{(2m_K^2+m_\pi^2)/3}=411$~MeV is the simulation centre-of-mass of the pseudoscalar meson octet. Ratios of moments are shown and the $X_\pi$ normalization is taken for the figures so that they may be easily compared against published results~\cite{Horsley:2010th,*Cloet:2012db}. The quality of fit is clearly acceptable in each case, with the $\chi^2/\textrm{dof}$ between 0.6 and 0.9 for each moment. All $\chi^2$ values are less than one as the effect of correlations between the original lattice data could not be included. Figures~\ref{fig:ZeroSD}, \ref{fig:FirstSD} and \ref{fig:FirstSI} show the fits to the data in a form suitable for the extraction of the CSV terms at the unphysical symmetric point by Eqs.~(\ref{eq:lineq}) and (\ref{eq:lineq2}). The full analysis, presented in the next section, includes an extrapolation to the physical point.

\begin{widetext}

\begin{figure}[ftbh]
\centering
\includegraphics[width=0.5\columnwidth]{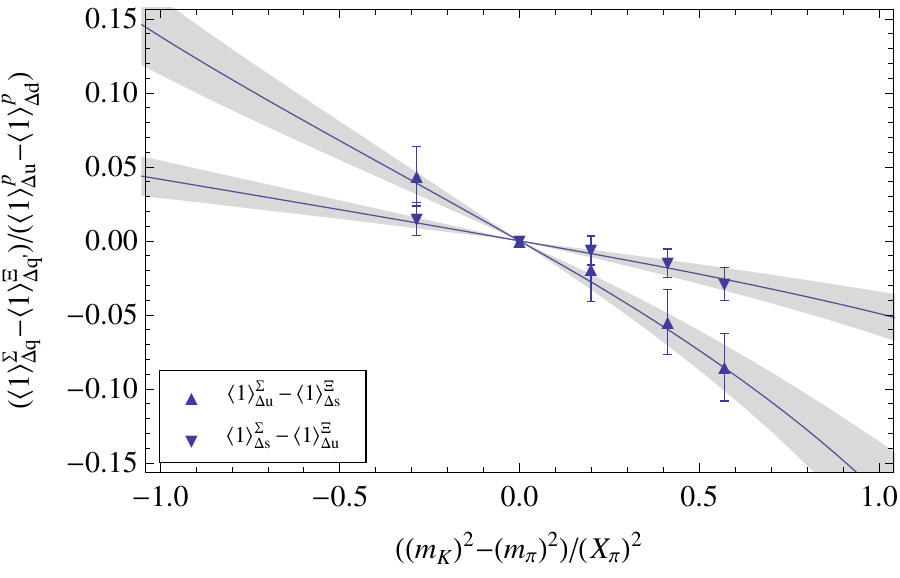}
\caption{Illustration of the fit to the zeroth spin-dependent moments -- data from Ref.~\cite{Horsley:2010th,Cloet:2012db}.}
\label{fig:ZeroSD}
\end{figure}
\begin{figure}[ftbh]
\centering
\includegraphics[width=0.5\columnwidth]{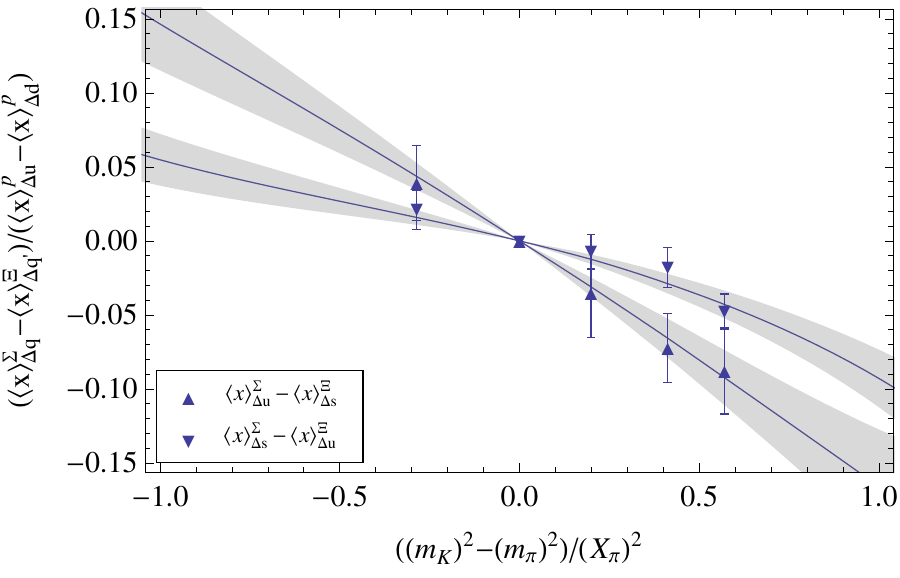}
\caption{Illustration of the fit to the first spin-dependent moments -- data from Ref.~\cite{Horsley:2010th,Cloet:2012db}.}
\label{fig:FirstSD}
\end{figure}
\begin{figure}[ftbh]
\centering
\includegraphics[width=0.5\columnwidth]{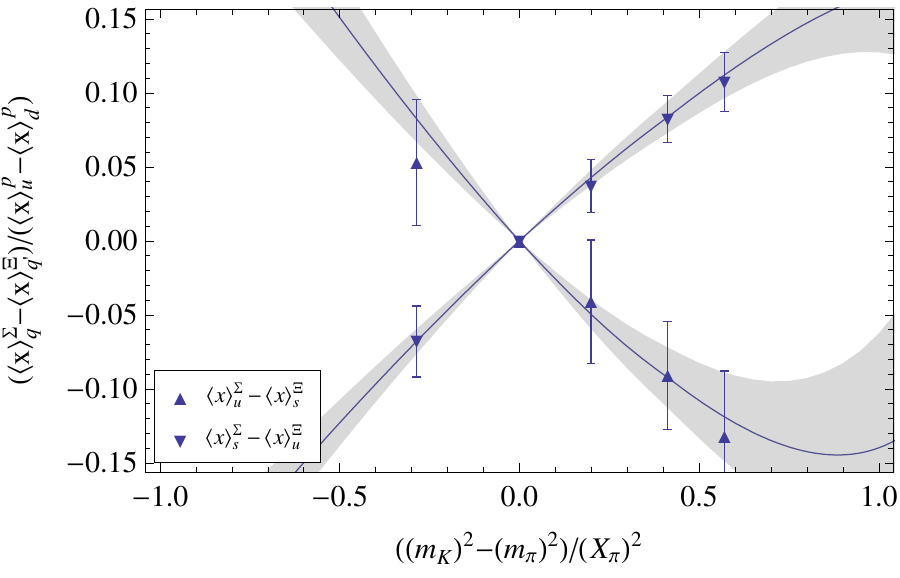}
\caption{Illustration of the fit to the first spin-independent moments -- data from Ref.~\cite{Horsley:2010th,Cloet:2012db}.}
\label{fig:FirstSI}
\end{figure}

\end{widetext}

\subsection{Evaluation of CSV terms}

As described in section~\ref{sec:method}, the CSV terms given in Eqs.~(\ref{eq:deltau}) and (\ref{eq:deltad}) may be evaluated by simply substituting the best-fit parameters of the isospin-averaged fit described in section~\ref{subsec:isoavfit} into the full SU(3), isospin-broken, perturbation theory expressions for the CSV terms. For example, $\delta \Delta u^m$ may be expressed as a function of quark mass in the form~\cite{Shanahan:2012}:
\begin{widetext}
\begin{equation}
\label{eq:deltaum}
\delta \Delta u^m = \langle x^m \rangle^p_{\Delta u} - \langle x^m \rangle^n_{\Delta d} =  a_{\Delta}^{(m)} + \frac{1}{16 \pi^2 f^2}\left(b_{\Delta}^{(m)} + d_{\Delta}^{(m)} + g_{\Delta}^{(m)}\right),
\end{equation}
where
\begin{align}
a_{\Delta}^{(m)} = & \frac{1}{2}\left(-\Delta n^{(m)}_1+\Delta n^{(m)}_2+\Delta n^{(m)}_3+\Delta n^{(m)}_6\right)B(m_u-m_d),\\[10pt] 
b_{\Delta}^{(m)} = & \frac{1}{6 \sqrt{3}}\left(D^2-2 D F-3 F^2\right) \sin (2 \epsilon ) \left(5 \Delta\alpha^{(m)} +2 \Delta\beta^{(m)} +6 \Delta\sigma^{(m)} \right)\left[\widetilde{J}(m_{\pi_0}^2)-\widetilde{J}(m_\eta^2)\right] \\ \nonumber
& {} + \frac{1}{24} \left[-D^2\left(9 \Delta\alpha^{(m)} +2 \Delta\beta^{(m)} +8 \Delta\sigma^{(m)} \right)+2 D F \left(19 \Delta\alpha^{(m)} +10 \Delta\beta^{(m)} + 24 \Delta\sigma^{(m)} \right)\right. \\ \nonumber
& {}~~~~~~~~~  \left. +3 F^2 \left(5 \Delta\alpha^{(m)} +2 \Delta\beta^{(m)} +8 \Delta\sigma^{(m)} \right)\right]\left[\widetilde{J}(m_{K^0}^2)-\widetilde{J}(m_{K^\pm}^2)\right] \\ \nonumber
& {}+ \frac{1}{24}\left(5\Delta \alpha^{(m)}+2\Delta \beta^{(m)}\right)\left[\widetilde{J}_T(m_{K^0}^2)-\widetilde{J}_T(m_{K^\pm}^2)\right],\\[10pt]
d_{\Delta}^{(m)} = & -\frac{1}{72}\left(5 \Delta\alpha^{(m)} + 2 \Delta\beta^{(m)} + 6 \Delta\sigma^{(m)}\right) \mathcal{C}^2\left[\widetilde{J}_2(m_{K^0}^2,\delta)-\widetilde{J}_2(m_{K^\pm}^2,\delta)\right]\\ \nonumber
 & -\frac{1}{108}\left(5\Delta\gamma^{(m)}-\Delta\gamma'^{(m)}\right) \mathcal{C}^2\left[\widetilde{J}_2(m_{K^0}^2,\delta)-\widetilde{J}_2(m_{K^\pm}^2,\delta)\right] \\[10pt]
g_{\Delta}^{(m)}=& {}-\frac{4}{9 \sqrt{3}} (D-3 F) \sin (2 \epsilon )\Delta\omega^{(m)}\left[\widetilde{J}_1(m_{\pi^0}^2, \delta)+\widetilde{J}_1(m_{\eta}^2, \delta)\right]\\ \nonumber
& {}+ \frac{2}{9}(D- 3F)\Delta\omega^{(m)}\left[\widetilde{J}_1(m_{K^0}^2,\delta)-\widetilde{J}_1(m_{K^\pm}^2,\delta)\right],
\end{align}
\end{widetext}
and expressions for the (subtracted) integrals $\widetilde{J}$ are given in the previous section. Clearly, entirely analogous expressions may be written for $\delta \Delta d^m$ and the spin-independent CSV terms. These, taken from Ref.~\cite{Shanahan:2012}, are given in appendix~\ref{app:CSVexpand}. We remind the reader that, to the same order in the broken SU(3) symmetry, analogous expressions for each quark flavour combination in each octet baryon are expressed in terms of different linear combinations of the same coefficients. The general case is given in Ref.~\cite{Shanahan:2012}.

In the above expression, meson masses take the form
\label{mesonmasses}
\begin{align}
m_{\pi^\pm}^2 = & B(m_u + m_d)\\ \nonumber
m_{\pi^0}^2  =& B(m_u+m_d) \\
&-\frac{2B}{3}(2m_s-(m_u+m_d))\frac{\textrm{sin}^2\epsilon}{\textrm{cos} 2 \epsilon}\\
m_{K^\pm}^2  =& B(m_s+m_u)\\
m_{K^0}^2 = & B(m_s+m_d)\\ \nonumber
m_{\eta}^2 = & \frac{B}{3}(4m_s+m_u+m_d)\\
& +\frac{2B}{3}(2m_s-(m_u+m_d))\frac{\textrm{sin}^2\epsilon}{\textrm{cos} 2 \epsilon},
\end{align}
and the $\pi^0-\eta$ mixing angle $\epsilon$ is given by
\begin{equation}
\textrm{tan} 2 \epsilon = \frac{\sqrt{3}~(m_d-m_u)}{2m_s-(m_d+m_u)}.
\end{equation}

The parameters $\Delta n_i^{(m)}$, $\Delta \alpha^{(m)}$, $\Delta \beta^{(m)}$ and $\Delta \sigma^{(m)}$ are determined, for $m=\{0,1\}$, from the isospin-averaged fits. All that remains to be specified for an evaluation of $\delta \Delta u^m$ from the expression above are values for $Bm_q$. 

To evaluate the CSV terms at the physical point we take as input the estimate for the physical up-down quark mass ratio from Ref.~\cite{Leutwyler:1996qg}
\begin{equation}
R:=\frac{m_u}{m_d} = 0.553 \pm 0.043,
\label{eq:R}
\end{equation}
determined by a fit to meson decay rates. We note that this value is compatible with more recent estimates of the ratio from $2+1$ and 3 flavor QCD and QED~\cite{Aubin2004,Blum2010}, and lies within uncertainties of the FLAG lattice averaging group estimate $R=0.47(4)$~\cite{Colangelo:2010et}.
The Gell-Mann-Oakes Renner relation suggests the definition
\begin{equation}
\omega = \frac{B(m_d-m_u)}{2} := \frac{1}{2} \frac{(1-R)}{(1+R)} m_{\pi_\textrm{\tiny{(phys)}}}^2,
\end{equation}
which allows one to define
\begin{align}
Bm_u & =m_{\pi_\textrm{\tiny{(phys)}}}^2/2 - \omega, \\
Bm_d & =m_{\pi_\textrm{\tiny{(phys)}}}^2/2 + \omega, \\
Bm_s & =m_{K_\textrm{\tiny{(phys)}}}^2-m_{\pi_\textrm{\tiny{(phys)}}}^2/2.
\end{align}
Here, $m_{\pi_\textrm{\tiny{(phys)}}}=137.3$~MeV and $m_{K_\textrm{\tiny{(phys)}}}=497.5$~MeV are taken to be the physical isospin-averaged meson masses~\cite{PDG}.

As the available QCDSF/UKQCD Collaboration lattice results are presented only in terms of ratios of moments, there is an unknown constant scaling factor $Z$ associated with all data points. This $Z$ is distinct for each moment (zeroth and first SD and first SI) under consideration. These constants are determined by matching the extrapolations for the isovector moments to experimental values at the physical point at 4~GeV$^2$~\cite{PDGold,Blumlein:2010rn,Martin:2002aw}:
\begin{align}
g_A=\langle 1 \rangle^p_{\Delta u - \Delta d} & \underset{\textrm{expt}}{=} 1.2695(29),\\
\langle x \rangle^p_{\Delta u - \Delta d} & \underset{\textrm{expt}}{=} 0.190(8),\\
\langle x \rangle^p_{ u - d} & \underset{\textrm{expt}}{=}0.157(9).
\end{align}

The uncertainty of these experimental numbers is propagated into the final results.
The full error analysis also takes account of correlated uncertainties between all of the fit parameters in the original fits~\cite{Shanahan:2012}, as well as allowing for the stated variation of $R$. The regulator mass $\Lambda=1$~GeV is allowed to vary by $\pm 20\%$, which is again propagated into the final uncertainty. Changing the regulator $u(k)$ within the FRR scheme leads to small variations of order $1\%$.

The results of this analysis are summarized and compared with previous work in Table~\ref{tab:results2}. While the light quark ratio $R$ was used as input in this calculation, the determination of the CSV terms via a linear flavour expansion~\cite{Horsley:2010th,Cloet:2012db} used the quark mass ratio ${3(m_d-m_u)}/{(m_d+m_u+m_s)}=0.066(7)$~\cite{Leutwyler:1996qg}. The choice of $R$ made here sets this ratio to the same value. The other inputs used in both calculations, namely the experimental isovector moments at the physical point, take the same values in both calculations. Thus, the linear and chiral results in Table~\ref{tab:results2} are directly comparable. 

In particular, evaluating the chiral perturbation theory expressions for the CSV terms at the point where $(m_d+m_u)=2m_s$ and both $(m_d-m_u)$ and $(m_u+m_d+m_s)$ take their physical values, labelled `SU(3)-sym' in Table~\ref{tab:results2}, gives results which may be directly compared with the linear flavor expansion calculation.
As might be anticipated from an inspection of Figs.~\ref{fig:ZeroSD}--\ref{fig:FirstSI} which show fits qualitatively consistent with straight lines, chiral loop corrections to the CSV terms at this point are small and within uncertainties. 

Comparison with results evaluated at the physical pseudoscalar masses gives an indication of the chiral loop corrections in moving away from the SU(3) point. Again, these corrections are small in the spin-dependent case, while being more significant in the spin-independent case. It is noted that, in contrast to the results of the linear flavour expansion, the chiral perturbation theory results are based on fits for the quark distribution moments to all lattice data simultaneously (for each moment), and thus include the proper correlations between quark moments in each of the baryons. As a consequence, even with more fit parameters, the uncertainties are comparable to the simple linear fits.

\begin{table*}[htb]
\centering
\begin{ruledtabular}
\begin{tabular}{l d d d}
 Moment & \multicolumn{1}{c}{Linear: SU(3)-sym} & \multicolumn{1}{c}{Chiral: SU(3)-sym}  & \multicolumn{1}{c}{Chiral: physical}\\
\hline
 $\delta \Delta u^{0+}$ & -0.0057(14) & -0.0063(13)  &  -0.0061(13)\\
 $\delta \Delta d^{0+}$ & -0.0018(6)  & -0.0019(6)   &  -0.0018(6) \\
 $\delta \Delta u^{1-}$ & -0.0010(3)  & -0.0007(2)   &  -0.0007(2) \\
 $\delta \Delta d^{1-}$ & -0.0004(1)  & -0.0003(1)   &  -0.0002(1) \\
 $\delta u^{1+}$        & -0.0012(3)  & -0.0013(3)   &  -0.0023(7)\\
 $\delta d^{1+}$        &  0.0010(2)  &  0.0012(2)   &   0.0017(4)\\
\end{tabular}
\end{ruledtabular}
\caption{Comparison of results. The column labelled `Linear' gives the results which were published with the lattice simulation results~\cite{Horsley:2010th,Cloet:2012db}, calculated using a linear flavor expansion about the SU(3) symmetric point. These have been corrected from the values quoted in~\cite{Horsley:2010th,Cloet:2012db}, as explained in the footnote preceding Eq.~(\ref{eq:lineq}). `Chiral' gives the results of this work, i.e., including chiral physics, both at the comparable `SU(3) symmetric' point (with $(m_d+m_u)=2m_s$ but the physical $(m_d-m_u)$), labelled `SU(3)-sym', and at physical pseudoscalar masses.}
\label{tab:results2}
\end{table*}

\begin{widetext}

\begin{figure}[tbhf]
\centering
\includegraphics[width=0.75\columnwidth]{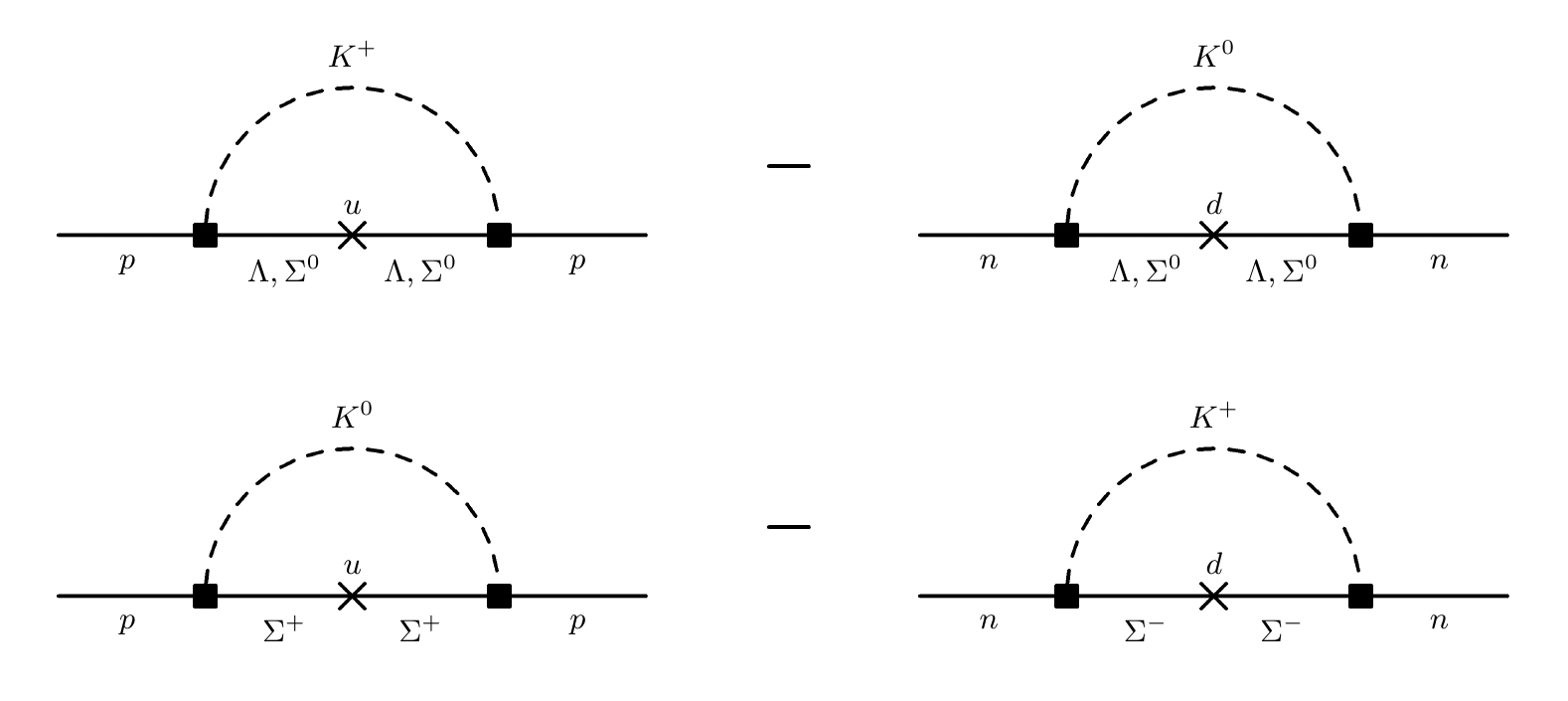}
\caption{Illustration of some of the octet loop terms contributing to $\delta \Delta u^m = \langle x^m \rangle^p_{\Delta u} - \langle x^m \rangle^n_{\Delta d}$ or $\delta u^m = \langle x^m \rangle^p_{u} - \langle x^m \rangle^n_{d}$. These contributions are non-vanishing when the loop pseudoscalar masses are different, i.e., when $m_{K^0}^2 \ne m_{K^\pm}^2$.}
\label{fig:kaonloops2}
\end{figure}

\end{widetext}

The origin of the chiral loop contributions to the CSV terms can be seen clearly from the form of Eq.~(\ref{eq:deltaum}) (and the analogous Eqs.~(\ref{eq:deltadm}), (\ref{eq:deltaUm}) and (\ref{eq:deltaDm}) in Appendix~\ref{app:CSVexpand}). One contribution to the $(u-d)$ moments is illustrated diagrammatically in Fig.~\ref{fig:kaonloops2}. The kaon loop diagrams shown, and the analogous diagrams for the $(d-u)$ moments, give contributions to the CSV terms proportional to $\left[ \widetilde{J}(m_{K^0}^2)-\widetilde{J}(m_{K^\pm}^2)\right]$, which is non-vanishing when $m_{K^0}^2\ne m_{K^\pm}^2$. The corresponding wavefunction renormalization terms, as well as tadpole and decuplet kaon-loop diagrams, also contribute to the CSV terms proportional to $\left[ \widetilde{J}(m_{K^0}^2)-\widetilde{J}(m_{K^\pm}^2)\right]$. In the spin-independent case, this kaon mass difference effect yields the only chiral loop corrections to the CSV terms. For the spin-dependent moments, however, additional terms proportional to $\left[ \widetilde{J}(m_{m_0}^2)-\widetilde{J}(m_{\eta}^2)\right]$ also contribute. Cancellation of octet loop terms with wavefunction renormalization contributions ensures that these terms vanish in the SI case.

The chiral loops also account for the corrections in moving from the `SU(3) point' to the physical point. For example, as one moves along the line of constant singlet quark mass ($(m_u+m_d+m_s)=\textrm{constant}$) from the SU(3) symmetric point to the physical point, the difference $\left[ \widetilde{J}(m_{K^0}^2)-\widetilde{J}(m_{K^\pm}^2)\right]$ decreases in magnitude by approximately 30\%.

%\begin{figure}[ftbh]
%\centering
%\includegraphics[width=\columnwidth]{ZeroSD.pdf}
%\caption{Illustration of the fit to the zeroth spin-dependent moments -- data from Ref.~\cite{Horsley:2010th,Cloet:2012db}.}
%\label{fig:ZeroSD}
%\end{figure}
%
%\begin{figure}[ftbh]
%\centering
%\includegraphics[width=\columnwidth]{FirstSD.pdf}
%\caption{Illustration of the fit to the first spin-dependent moments -- data from Ref.~\cite{Horsley:2010th,Cloet:2012db}.}
%\label{fig:FirstSD}
%\end{figure}
%
%\begin{figure}[ftbh]
%\centering
%\includegraphics[width=\columnwidth]{FirstSI.pdf}
%\caption{Illustration of the fit to the first spin-independent moments -- data from Ref.~\cite{Horsley:2010th,Cloet:2012db}.}
%\label{fig:FirstSI}
%\end{figure}

\section{Conclusion}

We have used a chiral perturbation theory analysis to extrapolate QCDSF/UKQCD Collaboration lattice data for the first several Mellin moments of quark distribution functions to the physical quark masses. This technique allows the charge symmetry violating (CSV) parton distributions to be evaluated at the physical point.

The conclusion of this study is quite clear. The chiral corrections to the spin-dependent CSV moments are very small.
%There are small chiral corrections to the CSV terms for both the spin-dependent and spin-independent moments.
In particular, an analysis of the same lattice data using a linear flavor expansion about the SU(3) symmetric point gave compatible results~\cite{Horsley:2010th,Cloet:2012db}. A detailed analysis shows that both the chiral corrections at the SU(3) symmetric point, as well as the extrapolation from this point to the physical quark or pseudoscalar masses, are small effects.

At the physical point, this analysis gives the spin-dependent CSV terms to be $\delta\Delta u^{0+}=-0.0061(13)$, $\delta\Delta d^{0+}=-0.0018(6)$, $\delta\Delta u^{1-}=-0.0007(2)$, and $\delta\Delta d^{1-}=-0.0002(1)$.
As a result, one would expect CSV corrections to the Bjorken sum rule~\cite{Bjorken:1966jh,Bjorken:1969mm} to appear at the half-percent level. Measuring these corrections would require significant improvement over the current best determination of the sum rule to $8\%$ precision at $Q^2=3~\textrm{GeV}^2$ from a recent COMPASS Collaboration experiment~\cite{Alekseev:2010hc}.

%\begin{align}
%&\int_0^1 dx \left[ g_1^p(x)-g_1^n(x) \right] = \frac{G_A}{6 G_V} \left[ 1 - \frac{\alpha_S(Q^2)}{\pi} \right] \\
%&= \left\langle \frac{\Delta u(x) - \Delta d(x)}{6}+ \frac{4 \delta \Delta d(x) + \delta \Delta u(x)}{18} \right\rangle
%\end{align}

For the spin-independent moments, the chiral corrections are more significant. This analysis gives $\delta u^{1+}=-0.0023(7)$ and $\delta d^{1+}=0.0017(4)$, in good agreement with previous phenomenological estimates of CSV both within the MIT bag model~\cite{Rodionov:1994cg,Londergan:2003pq} and using the MRST analysis~\cite{Martin:2003sk}. These results support the conclusion~\cite{Londergan:2003pq,Londergan:2009kj} that partonic CSV effects may reduce the $3\sigma$ discrepancy with the standard model reported by the NuTeV Collaboration~\cite{Zeller:2001hh} by up to $30 \%$.

\section*{Acknowledgements}
We gratefully acknowledge the assistance of the QCDSF/UKQCD Collaboration in providing access to the raw lattice data. We also acknowledge helpful discussions with J.~Zanotti.
This work was supported by the University of Adelaide and the Australian
Research Council through through the ARC Centre of Excellence for Particle Physics at the Terascale and grants FL0992247 (AWT) and DP110101265 (RDY) and FT120100821 (RDY).

\appendix

\begin{widetext}

\section{Extrapolation formulae}
\label{app:CSVexpand}

This section gives formulae for the spin-dependent and spin-independent charge symmetry violating quark distributions as functions of quark and meson mass. All integrals are defined in the body of the report.

The fit parameters which appear in the following expressions are discussed in detail in Ref.~\cite{Shanahan:2012}. For the zeroth spin-dependent moment there are eight free parameters; six linearly independent linear coefficients $\Delta n_{i=1-6}^{(0)}$, the baryon-baryon-meson coupling constant $D$, and an operator insertion parameter $\Delta \sigma^{(0)}$. 
Both the first spin-dependent and first spin-independent moment have nine fit parameters; six linear coefficients $\Delta n_{i=1-6}^{(1)}$ ($ n_{i=1-6}^{(2)}$), and three operator insertion parameters $\Delta \alpha^{(1)}$, $\Delta \beta^{(1)}, \Delta \sigma^{(1)}$ ($\alpha^{(2)}$, $\beta^{(2)}, \sigma^{(2)}$) in the spin-dependent (-independent) case.

Baryon-baron-meson couplings $F$ and $D$ are set to their physical values by $D \rightarrow D_{\textrm{phys}}=\frac{3}{5}g_{A_{\textrm{phys}}}$ for each of the first-moment fits.  
For all three fits, SU(6) symmetry is used to set $F=\frac{2}{3}D$ and $\mathcal{C}\rightarrow \mathcal{C}_{\textrm{phys}} = -\frac{6}{5}g_{A_{\textrm{phys}}}$ is fixed. Decuplet ($\gamma$) and transition ($\omega$) insertion parameters are also fixed for each fit, either by using SU(6) symmetry to relate them to other fit parameters, or, in the case of $\gamma$ for the first spin-independent moment, by using an experimental result, as detailed in Ref.~\cite{Shanahan:2012}. 

\subsection{Spin-dependent CSV terms}

This section gives an explicit expression for the spin-dependent CSV distribution $\delta \Delta d^m$ as a function of quark and meson mass. The corresponding expression for $\delta \Delta u^m$ is given in the body of the report.

\begin{equation}
\label{eq:deltadm}
\delta \Delta d^m = \langle x^m \rangle^p_{\Delta d} - \langle x^m \rangle^n_{\Delta u} =  \overline{a}_{\Delta}^{(m)} + \frac{1}{16 \pi^2 f^2}\left(\overline{b}_{\Delta}^{(m)} + \overline{d}_{\Delta}^{(m)} +\overline{g}_{\Delta}^{(m)}\right)
\end{equation}
\begin{align}
\overline{a}_{\Delta}^{(m)} = & \frac{1}{2}\left(-\Delta n^{(m)}_3+\Delta n^{(m)}_6\right)B(m_u-m_d)\\[10pt] 
\overline{b}_{\Delta}^{(m)} = & \frac{1}{6 \sqrt{3}}\left(D^2-2 D F-3 F^2\right) \sin (2 \epsilon ) \left(\Delta\alpha^{(m)} +4 \Delta\beta^{(m)} +6 \Delta\sigma^{(m)} \right)\left[\widetilde{J}(m_{\pi_0}^2)-\widetilde{J}(m_\eta^2)\right] \\ \nonumber
& {} +\frac{1}{24} \left[D^2 \left(\Delta\alpha^{(m)} -4 \Delta\beta^{(m)} -8 \Delta\sigma^{(m)} \right)+6 D F \left(\Delta\alpha^{(m)} +4 \Delta\beta^{(m)} +8 \Delta\sigma^{(m)} \right)\right. \\ \nonumber
& {}~~~~~~~~~ \left. +F^2 \left(5 \Delta\alpha^{(m)} +20 \Delta\beta^{(m)} +24 \Delta\sigma^{(m)} \right)\right]\left[\widetilde{J}(m_{K^0}^2)-\widetilde{J}(m_{K^\pm}^2)\right] \\ \nonumber
& {}- \frac{1}{24}\left(\Delta \alpha^{(m)}+4\Delta \beta^{(m)}\right)\left[\widetilde{J}_T(m_{K^0}^2)-\widetilde{J}_T(m_{K^\pm}^2)\right]\\[10pt]
\overline{d}_{\Delta}^{(m)} = & -\frac{1}{72}\left(\Delta\alpha^{(m)} + 4 \Delta\beta^{(m)} + 6 \Delta\sigma^{(m)}\right)\mathcal{C}^2\left[\widetilde{J}_2(m_{K^0}^2,\delta)-\widetilde{J}_2(m_{K^\pm}^2,\delta)\right]\\ \nonumber
 & +\frac{1}{324}\left(5\Delta\gamma^{(m)}-\Delta\gamma'^{(m)}\right)\mathcal{C}^2\left[\widetilde{J}_2(m_{K^0}^2,\delta)-\widetilde{J}_2(m_{K^\pm}^2,\delta)\right] \\[10pt]
\overline{g}_{\Delta}^{(m)} = & {}+\frac{4}{9 \sqrt{3}} (D-3 F) \sin (2 \epsilon )\Delta\omega^{(m)}\left[\widetilde{J}_1(m_{\pi^0}^2, \delta)+\widetilde{J}_1(m_{\eta}^2, \delta)\right]\\ \nonumber
& {}+ \frac{4}{9}F\Delta\omega^{(m)}\left[\widetilde{J}_1(m_{K^0}^2,\delta)-\widetilde{J}_1(m_{K^\pm}^2,\delta)\right].
\end{align}

\subsection{Spin-independent CSV terms}

This section gives explicit expressions for the spin-independent CSV distributions as functions of quark and meson mass.

\begin{align}
\label{eq:deltaUm}
\delta u^m = &\langle x^m \rangle^p_{u} - \langle x^m \rangle^n_{d} =  a^{(m)} + \frac{1}{16 \pi^2 f^2}\left(b^{(m)} + d^{(m)}\right)\\ \label{eq:deltaDm}
\delta  d^m = &\langle x^m \rangle^p_{d} - \langle x^m \rangle^n_{u} =  \overline{a}^{(m)} + \frac{1}{16 \pi^2 f^2}\left(\overline{b}^{(m)} + \overline{d}^{(m)} \right)
\end{align}
\begin{align}
a^{(m)} = & \frac{1}{2}\left(- n^{(m)}_1+ n^{(m)}_2+ n^{(m)}_3+ n^{(m)}_6\right)B(m_u-m_d)\\[10pt] 
b^{(m)} = & \frac{1}{24} \left[D^2\left(7 \alpha^{(m)} -2 \beta^{(m)}\right)+6 D F \left(\alpha^{(m)} -2\beta^{(m)}\right) +3 F^2 \left(5 \alpha^{(m)} +2 \beta^{(m)} \right)\right]\left[\widetilde{J}(m_{K^0}^2)-\widetilde{J}(m_{K^\pm}^2)\right] \\ \nonumber
& {}+ \frac{1}{24}\left(5 \alpha^{(m)}+2 \beta^{(m)}\right)\left[\widetilde{J}_T(m_{K^0}^2)-\widetilde{J}_T(m_{K^\pm}^2)\right]\\[10pt]
d^{(m)} = & -\frac{1}{72}\left(5 \alpha^{(m)} + 2 \beta^{(m)} + 6 \sigma^{(m)}\right)\mathcal{C}^2\left[\widetilde{J}_2(m_{K^0}^2,\delta)-\widetilde{J}_2(m_{K^\pm}^2,\delta)\right]\\ \nonumber
 & -\frac{1}{36}\left(3\gamma^{(m)}-\gamma'^{(m)}\right)\mathcal{C}^2\left[\widetilde{J}_2(m_{K^0}^2,\delta)-\widetilde{J}_2(m_{K^\pm}^2,\delta)\right].\\[20pt]
\overline{a}^{(m)} = & \frac{1}{2}\left(- n^{(m)}_3+ n^{(m)}_6\right)B(m_u-m_d)\\[10pt] 
\overline{b}^{(m)} = & \frac{1}{24} \left[-D^2 \left(7\alpha^{(m)} +4 \beta^{(m)} \right)+6 D F \left(\alpha^{(m)} +4 \beta^{(m)} \right)-3F^2 \left(\alpha^{(m)} +4 \beta^{(m)} \right)\right]\left[\widetilde{J}(m_{K^0}^2)-\widetilde{J}(m_{K^\pm}^2)\right] \\ \nonumber
& {}- \frac{1}{24}\left( \alpha^{(m)}+4 \beta^{(m)}\right)\left[\widetilde{J}_T(m_{K^0}^2)-\widetilde{J}_T(m_{K^\pm}^2)\right]\\[10pt]
\overline{d}^{(m)} = & -\frac{1}{72}\left(\alpha^{(m)} + 4 \beta^{(m)} + 6 \sigma^{(m)}\right)\mathcal{C}^2\left[\widetilde{J}_2(m_{K^0}^2,\delta)-\widetilde{J}_2(m_{K^\pm}^2,\delta)\right]\\ \nonumber
 & +\frac{1}{108}\left(3\gamma^{(m)}-\gamma'^{(m)}\right)\mathcal{C}^2\left[\widetilde{J}_2(m_{K^0}^2,\delta)-\widetilde{J}_2(m_{K^\pm}^2,\delta)\right].
\end{align}

\end{widetext}

\section{Figures}
\label{app:figs}

This section shows the fits to QCDSF/UKQCD lattice results discussed in section~\ref{subsec:isoavfit}. The figures are taken from Ref.~\cite{Shanahan:2012}, and are included here to give an indication of the quality of fit.

\begin{figure}[fbt]
\centering
\subfigure[Ratio of singly-represented quark moments.]{
\includegraphics[width=\columnwidth]{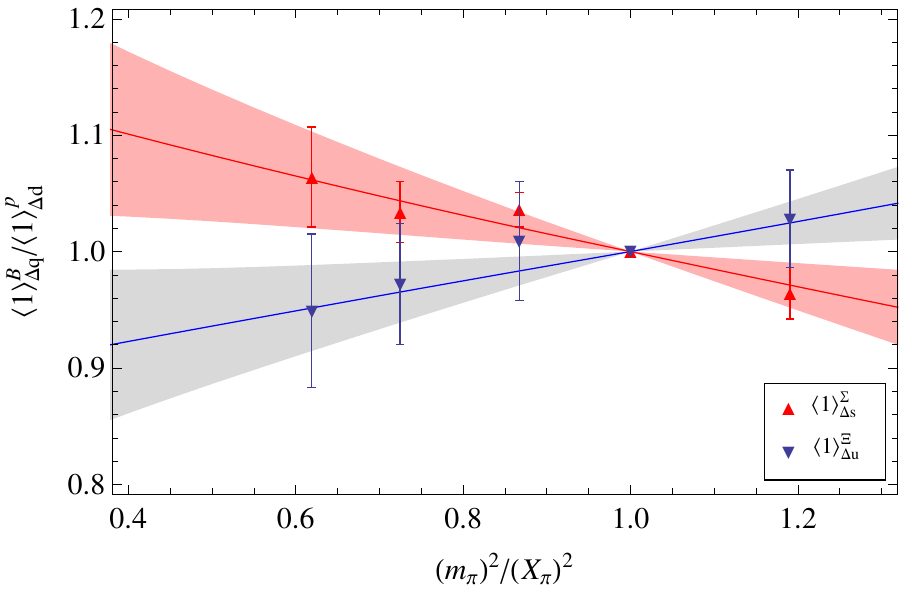}
\label{fig:ZeroSDSingly}
}
\subfigure[Ratio of doubly-represented quark moments.]{
\centering
\includegraphics[width=\columnwidth]{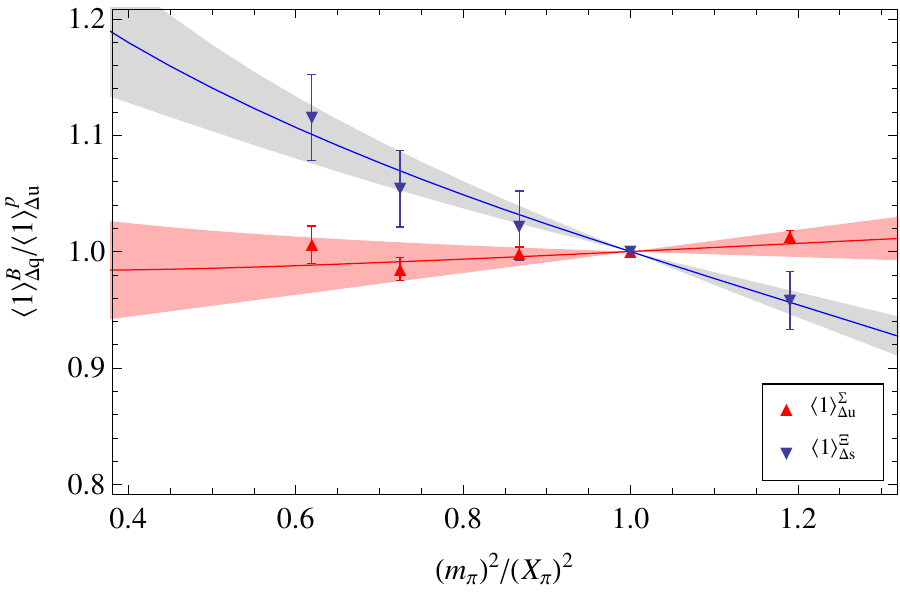}
\label{fig:ZeroSDDoubly}
}
\caption{Illustration of the fit to the zeroth spin-dependent moments -- data from Ref.~\cite{Horsley:2010th,Cloet:2012db}.}
\label{fig:SD0}
\end{figure}

\begin{figure}[fbt]
\centering
\subfigure[Ratio of singly-represented quark moments.]{
\includegraphics[width=\columnwidth]{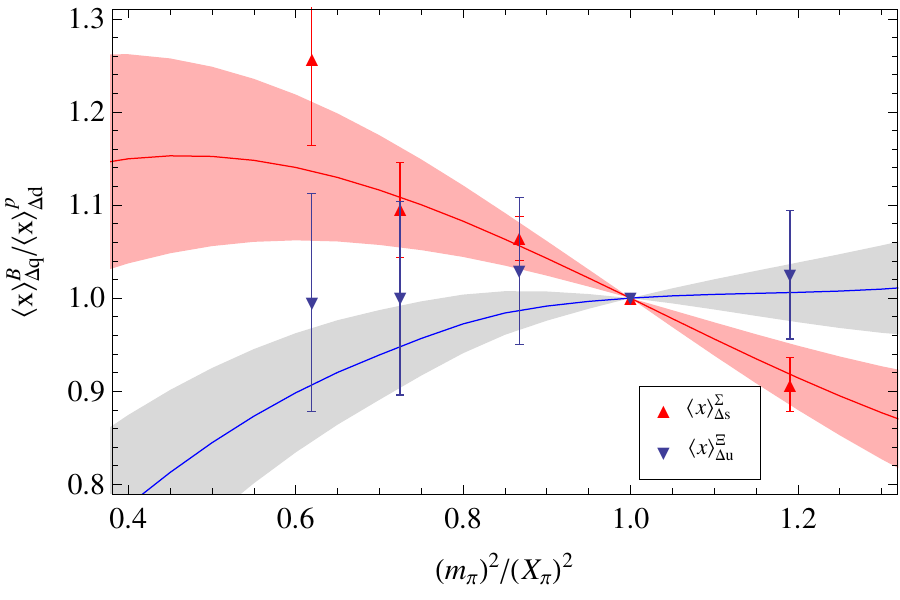}
\label{fig:FirstSDSingly}
}
\subfigure[Ratio of doubly-represented quark moments.]{
\centering
\includegraphics[width=\columnwidth]{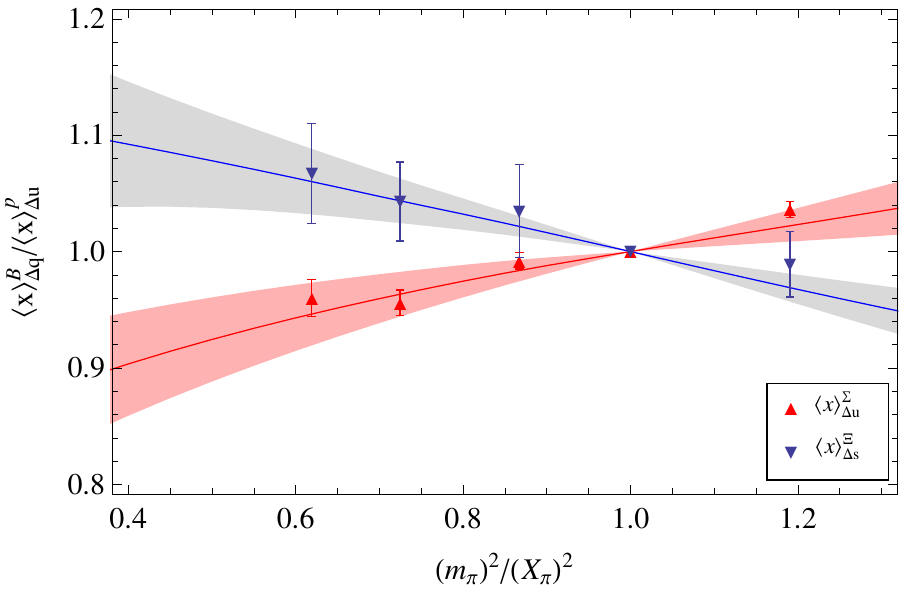}
\label{fig:FirstSDDoubly}
}
\caption{Illustration of the fit to the first spin-dependent moments -- data from Ref.~\cite{Horsley:2010th,Cloet:2012db}.}
\label{fig:SD1}
\end{figure}

\begin{figure}[fbt]
\centering
\subfigure[Ratio of singly-represented quark moments.]{
\includegraphics[width=\columnwidth]{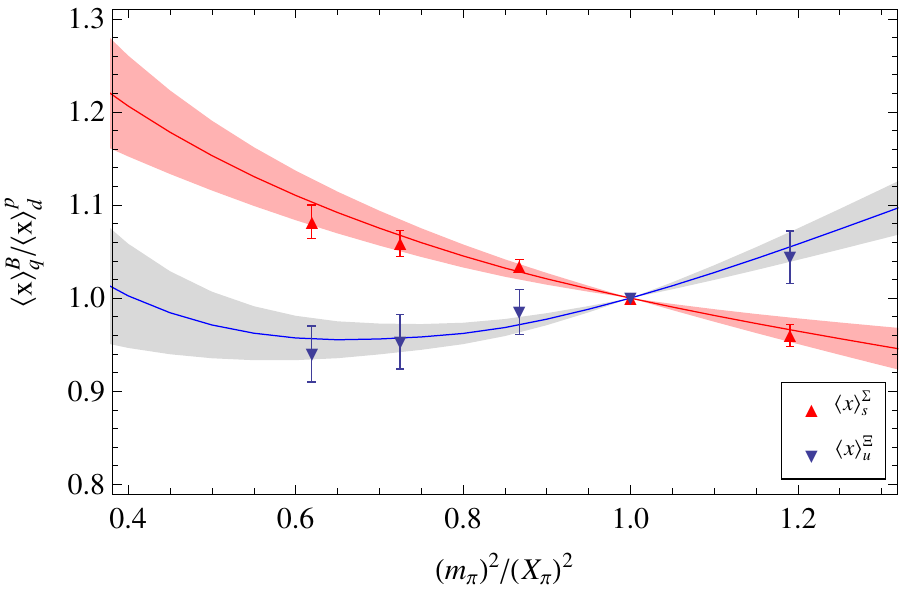}
\label{fig:FristSISingly}
}
\subfigure[Ratio of doubly-represented quark moments.]{
\centering
\includegraphics[width=\columnwidth]{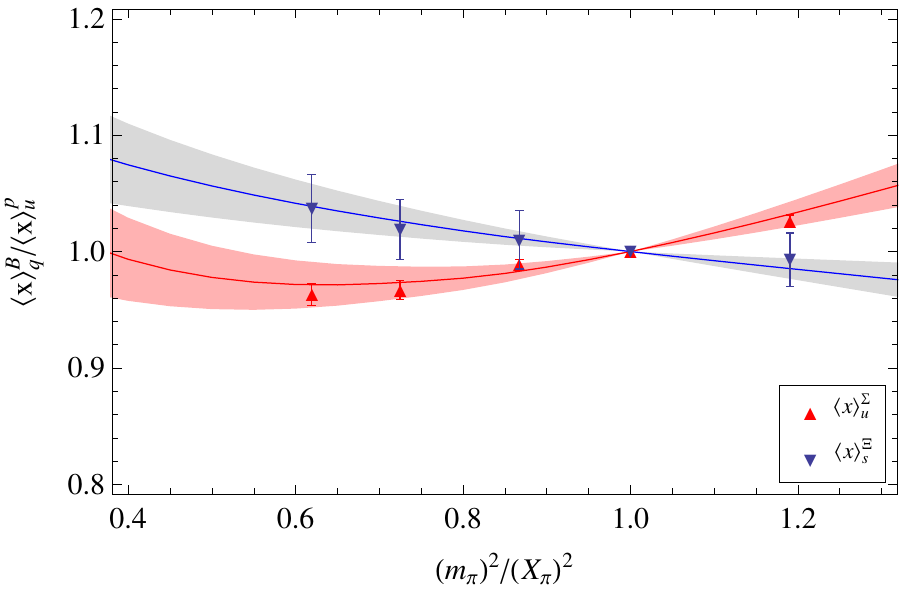}
\label{fig:FirstSIDoubly}
}
\caption{Illustration of the fit to the first spin-independent moments -- data from Ref.~\cite{Horsley:2010th,Cloet:2012db}.}
\label{fig:SI1}
\end{figure}

\end{document}